\documentclass[useAMS,usenatbib]{mn2e}

\usepackage{epsfig}

\def\gsim{\;\rlap{\lower 2.5pt
 \hbox{$\sim$}}\raise 1.5pt\hbox{$>$}\;}

\def\lsim{\;\rlap{\lower 2.5pt
   \hbox{$\sim$}}\raise 1.5pt\hbox{$<$}\;}

\def\ie{{\it i.e. }}

\def\eg{{\it e.g. }}

\def\dS{\Delta S}

\def\dfdb2{\partial_{b_2}f(s)}

\title[Halo Merger Rate in Ellipsoidal Collapse Model]{Conditional Mass
  Functions and Merger Rates of Dark Matter Halos in the Ellipsoidal
  Collapse Model}

\author[Jun Zhang, Chung-Pei Ma and Onsi Fakhouri]
{Jun Zhang$^{1}$\thanks{E-mail:jzhang@astro.berkeley.edu}, Chung-Pei Ma$^{1}$, Onsi Fakhouri$^{1}$\\
  \\
  $^{1}$601 Campbell Hall, Department of Astronomy, University of California, Berkeley, CA 94720, USA\\
}

\begin{document}

\pagerange{\pageref{firstpage}--\pageref{lastpage}} \pubyear{2006}
\maketitle
\label{firstpage}

\begin{abstract}
  Analytic models based on spherical and ellipsoidal gravitational collapse
  have been used to derive the mass functions of dark matter halos and
  their progenitors (the {\it conditional} mass function).  The ellipsoidal
  model generally provides a better match to simulation results, but there
  has been no simple analytic expression in this model for the conditional
  mass function that is accurate for small time steps, a limit that is
  important for generating halo merger trees and computing halo merger
  rates.  We remedy the situation by deriving accurate analytic formulae
  for the first-crossing distribution, the conditional mass function, and
  the halo merger rate in the ellipsoidal collapse model in the limit of
  small look-back times.  We show that our formulae provide a closer match
  to the Millennium simulation results than those in the spherical collapse
  model and the ellipsoidal model of Sheth \& Tormen (2002).
\end{abstract}


\section{Introduction}
\label{intro}

\cite{PS74} presented an analytical expression for the unconditional mass
function of dark matter halos at redshift $z$, $n(M,z)$, based on the
spherical collapse model.  This function is closely related to the
first-crossing distribution of random walks with a barrier in the excursion
set framework \citep{BCEK91}.  In this framework, the linear over-density
computed at a given point in the Lagrangian space fluctuates as a Markovian
process when smoothed on successively smaller scales, and a dark matter
halo is identified at the point when the random walk of the linear
over-density crosses a critical value, or a barrier $\mathcal{B}(M,z)$.  In
the spherical collapse model, this barrier depends only on time and is
independent of mass: $\mathcal{B}=\delta_c/D(z)$, where $\delta_c=1.68$ and
$D(z)$ is the standard linear growth factor.

Although analytically simple, the spherical collapse model has been found
to over-predict the abundance of small halos and under-predict that of
massive ones (\eg, \citealt{LC94,GB94,T98,ST99}). The reason is
mainly that halo collapses are generally triaxial rather than spherical
(\eg, \citealt{d70,bbks86}).  Based on \cite{BM96}, \cite{SMT01} use the
ellipsoidal collapse model and obtain fitting functions that provide a
closer match to the unconditional halo mass function in N-body simulations.
Unlike the spherical collapse model in which the condition for the
virialization of a dark matter halo is solely determined by the linear
over-density on the scale of the halo mass, the virialization condition in
the ellipsoidal collapse model also depends on halo ellipticity and
prolateness.  By assuming that a dark matter halo becomes virialized when
its third axis collapses, \cite{SMT01} find a new criterion for the
virialization of dark matter halos, which involves all three
parameters. They further simplify the virialization condition by fixing the
ellipticity and the prolateness at their most likely values for a given
over-density, and obtain a fitting formula for the barrier
$\mathcal{B}(M,z)$ that is mass-dependent, in contrast to the constant
$\mathcal{B}(z)$ of the spherical collapse model.  A mass-dependent barrier
is commonly referred to as a {\it moving} barrier, and it is this
mass-dependence that suppresses the abundance of small halos while
increasing that of massive ones in the ellipsoidal collapse model.
Physically, this is because a smaller halo typically has a larger
ellipticity and therefore a longer collapsing time.

The relationship between the unconditional mass function and the first
crossing distribution associated with barrier-crossing random walks has
been extended to obtain the {\it conditional} mass function of halos
\citep{BCEK91, LC93}.  In this so-called extended Press-Schechter (EPS)
model, the conditional mass function $dN(M_1,z_1|M_0, z_0)/dM_1$ gives the
average number of progenitor halos (of mass $M_1$ at redshift $z_1$) per
unit mass associated with a descendant halo of mass $M_0$ at redshift $z_0$
$\left(z_1>z_0\right)$.  Once determined, it can be used to generate merger
trees of halos for many applications (e.g., galaxy formation, growth of the
central black hole, reionization) using Monte Carlo simulations.

The conditional mass function has a simple analytic form in the constant
barrier spherical collapse model \citep{LC93}.  For a moving barrier (such
as the ellipsoidal collapse model), however, exact analytic forms have been
found only for the special case of a linear barrier (\citealt{ST02}, ST02
hereafter); while the same authors have proposed a Taylor-series-like
approximation for a general moving barrier.  We find that none of these
formulae work well for $z_1-z_0 \ll 1$, which is important for generating accurate merger trees in most Monte Carlo methods (\eg, \citealt{LC93,KW93,SK99,SL99,CLBF00}) and for relating halo merger rates to the conditional mass function (\S~2 below).  \cite{FM07} compare the halo merger rates determined from the Millennium simulation
(\citealt{springel05}) with the prediction of the spherical collapse EPS model, finding the latter to overpredict the major merger rates by up to a
factor of $\sim 2$ and underpredict the minor merger rates by up to a
factor of $\sim 5$.  Recently, various other fitting forms for the
conditional mass function have been proposed that are calibrated to the
results from particular $N$-body simulations, e.g., \cite{Cole07,
  Parkinson07,ND07}.  An alternative way of deriving the conditional mass
function that does not rely explicitly on fitting to simulations is to
solve the integral equation proposed by \cite{ZH06} (ZH06 hereafter). This
method, which is based on the conservation of probability in the excursion
set formalism, is accurate but computationally expensive.

In \S2, we derive almost exact analytic forms for the first crossing
distribution, the conditional mass function, and the halo merger rate in
the ellipsoidal collapse model in the limit of small look-back times; a
limit where earlier work breaks down.  Our method is based on ZH06, but our
results are expressed in simple analytic forms.  We compare the predictions
of this improved ellipsoidal collapse model with those from the spherical
collapse model and the Millennium simulation in \S\ref{Millennium}.  We
assume the cosmological parameters used in the Millennium simulation:
$\Omega_m=0.25$, $\Omega_b=0.045$, $h=0.73$, $\Omega_{\Lambda}=0.75$,
$n=1$, $\sigma_8=0.9$.

\section{Improved Ellipsoidal Collapse Model for Small Time Steps}
\label{EEPS}

We use $dN(M_1,z_1|M_0, z_0)/dM_1$ to denote the conditional mass function
of dark matter halos, defined in \S1.  We use $B_M(M_1,M_2,z)dM_1dM_2$ to
denote the merger rate of halos, which is defined to be the number of
mergers between halos of mass $(M_1, M_1+dM_1)$ and $(M_2, M_2+dM_2)$ per
unit volume and unit redshift at redshift $z$.  If halo mergers are assumed
to be binary and mass conserving such that the consequence of each merger
is the formation of a descendant halo of mass $M_0=M_1+M_2$, then we can
relate the merger rate to the (number-weighted) conditional mass function
through a simple relation (see \citealt{FM07} for a detailed discussion, 
or \citealt{SP97} for an earlier discussion):
\begin{eqnarray}
\label{relation}
    &&B_M(M_1,M_2,z)dM_1dM_2
    =n(M_1+M_2,z)d(M_1+M_2)\nonumber\\
    &\times&\frac{1}{\Delta z}\frac{d}{dM_1}N(M_1, z+\Delta z|M_1+M_2, z)dM_1
\end{eqnarray}  
where $n(M,z)$ is the (unconditional) halo mass function at redshift $z$,
and $\Delta z$ is assumed to be small.  

To study how many progenitors at $z_1$ are associated with a
descendant halo of mass $M_0$ at $z_0$, we let the random walk of the
linear over-density start from the scale of the descendant halo $S(M_0)$
with an over-density of $\mathcal{B}[S(M_0),z_0]$, where $S(M)=\sigma^2(M)$
is the variance of the linear density field smoothed with a window function
containing mass $M$.  A progenitor of mass $M_1$ is then identified once
the random walk crosses $\mathcal{B}[S(M_1),z_1]$ on the scale of $S(M_1)$.
The conditional mass function can then be written as
\begin{equation}
\label{dNdM_f}
            M_1\frac{d}{dM_1}N(M_1, z_1|M_0, z_0)dM_1=M_0f(\Delta S)d\Delta S
\end{equation}
where $\Delta S=S(M_1)-S(M_0)$, and $f(\Delta S)$ is the first-crossing
distribution of random walks with a barrier of the form $b(\Delta
S)=\mathcal{B}[S(M_1),z_1]-\mathcal{B}[S(M_0),z_0]$.

As mentioned in \S~1, analytic solutions for $f(\Delta S)$ have only been
found when the barrier is a constant or a linear function of $\Delta S$.
Here we propose an almost exact analytic solution for the first-crossing
distribution in the limit of $z_1-z_0 \ll 1$.  We use the fact that the
barrier $b(\Delta S)$ is a weakly nonlinear function of $\Delta S$ to approximate it by
\begin{equation}
\label{bds}
  b(\dS) = b_0+b_1 \dS + b_2(\dS)^2  \,,
\end{equation}
where $b_2 (\Delta S)^2$ is assumed to be subdominant in comparison with
the other two terms\footnote{This is obviously not true when $\Delta S$ is
  very large. But we find that in practice, the relevant range of $\Delta
  S$ is rarely large enough to invalidate the assumption.}.  The
first-crossing distribution can then be written as $f(\Delta
S,b_0,b_1,b_2)$.  When $b_2=0$, $f(\Delta S,b_0,b_1,0)$ has the analytic
form (see ST02):
\begin{equation}
\label{flinear}
     f(\Delta S,b_0,b_1,0)=\frac{b_0}{\Delta S\sqrt{2\pi \Delta S}}
     \exp\left[-\frac{(b_0+b_1\Delta S)^2}{2\Delta S}\right]
\end{equation}
For nonzero $b_2$, we approximate $f(\Delta S,b_0,b_1,b_2)$ as
\begin{eqnarray}
\label{approxi1}
   f(\Delta S,b_0,b_1,b_2) &\approx & f(\Delta S,b_0,b_1,0)\\ \nonumber 
   &+&b_2\times\left. \partial_{b_2} f(\Delta S,b_0,b_1,b_2) \right\vert_{b_2=0}
\end{eqnarray}
and derive the second term on the right side of eq.~(\ref{approxi1}) next.
  
For notational simplicity, we denote $\Delta S$ as $s$ and
$f(s,b_0,b_1,b_2)$ as $f(s)$.  The first-crossing $f(s)$ in general
satisfies the integral equation (see ZH06 for the derivation):
\begin{equation}
\label{f}
   f(s)=g_1(s)+\int_0^s ds' f(s') g_2(s,s')
\end{equation}
where 
\begin{eqnarray}
\label{g1g2p0}
  g_1(s)&=&\left[\frac{b(s)}{s}-2\frac{db}{ds}\right]P_0\left[b(s),s\right]\\ \nonumber
  g_2(s,s')&=&\left[2\frac{db}{ds}-\frac{b(s)-b(s')}{s-s'}\right]P_0
      \left[b(s)-b(s'),s-s'\right]\\ \nonumber
  P_0(\delta,s)&=&\frac{1}{\sqrt{2\pi s}}\exp \left(-\frac{\delta^2}{2s}\right)
\end{eqnarray}
From eq.~(\ref{f}) and after some algebra, we find
\begin{equation}
\label{dfdb2_2}
   \partial_{b_2}f(s)\vert_{b_2=0} = g(s) + \int_0^sds' \partial_{b_2} f(s') g_2(s,s') \vert_{b_2=0} \,,
\end{equation}
where
\begin{eqnarray}
\label{g1prime}
&&  g(s) = -b_0^2\times P_0(b_0+b_1s,s)\\ \nonumber
&+& \int_0^sds'f(s',b_0,b_1,0)(b_1^2s'^2-s')P_0\left[b_1(s-s'),s-s'\right] 
\end{eqnarray}

\begin{equation}
\label{g2prime}
g_2(s,s')\vert_{b_2=0} = b_1\times P_0\left[b_1(s-s'),s-s'\right] \, .
\end{equation}
The two terms in $g(s)$ come from $\partial_{b_2}g_1$ and
$\partial_{b_2}g_2$ respectively, and a number of terms have been cancelled out
using the relation $\int_0^sds'f(s',b_0,b_1,0) P_0[b_1(s-s'),s-s']=
P_0((b_0+b_1s,s)$ that follows from eq.~(\ref{f}) for $b_2=0$.

The complicated form of $g(s)$ in eq.~(\ref{g1prime}) makes it difficult to
solve eq.~(\ref{dfdb2_2}).  However, we are interested in the limit of
small look-back times for the conditional mass function, which corresponds
to a small barrier difference $b(s)$.  We can thus neglect the terms of
${\cal{O}}(b_0^2)$ and simplify $g$ as\footnote{Note that
  eq.(\ref{g1_analytic}) is derived by neglecting the first term on the
  right side of eq.(\ref{g1prime}) and approximating $f(s',b_0,b_1,0)$
  (eq.(\ref{flinear})) as $b_0/\sqrt{2\pi s^3}\times\exp(-b_1^2s/2)$. These
  approximations introduce errors of order $b_0^2$ to $g(s\gg b_0^2)$ and
  of order $b_0$ to $g(s\sim b_0^2)$. Similarly, the error on
  $\partial_{b_2}f(s)$ is of order $b_0^2$ when $s\gg b_0^2$, and of order
  $b_0$ when $s\sim b_0^2$. We find that this error is negligible when
  $b_0$ is small.}:
\begin{equation}
\label{g1_analytic}
   g(s)\approx\frac{b_0}{4}(b_1^2s-2)\exp\left(-\frac{b_1^2s}{2}\right) \,.
\end{equation}
We then solve for $\partial_{b_2}f(s)$ by combining eqs.~(\ref{dfdb2_2}),
(\ref{g2prime}), and (\ref{g1_analytic}) and using the Laplace transform:
\begin{equation}
\label{df_result}
     \partial_{b_2}f(s) |_{b_2=0}=-\frac{b_0}{4}\exp\left(-\frac{b_1^2s}{2}\right)
       \left[1+\frac{b_1\sqrt{s}}{\Gamma(3/2)}\right]  \,.
\end{equation}
Substituting this expression back into eq.~(\ref{approxi1}), we obtain
the first-crossing distribution for a barrier of the form
$b(\dS)=b_0+b_1 \dS + b_2 (\dS)^2$:
\begin{eqnarray}
\label{f_complete}
    f(\dS) &=& \frac{b_0}{\dS\sqrt{2\pi \dS}}\exp\left[-\frac{(b_0+b_1\dS)^2}{2\dS}\right]\\ 
    \nonumber
 &-&\frac{b_0b_2}{4}\exp\left(-\frac{b_1^2\dS}{2}\right)\left[1+\frac{b_1\sqrt{\dS}}
   {\Gamma(3/2)}\right]  + {\cal{O}}(b_0^2) \,.
\end{eqnarray}
This is our main result.  We note that this equation reduces to the
analytic expressions of eq.~(\ref{flinear}) in the spherical collapse model
(i.e. a constant barrier with $b_1=b_2=0$) and the ellipsoidal model with a
linear barrier (i.e. $b_2=0$).  The second term on the right hand side of
eq.~(\ref{f_complete}) is a new term arising from the quadratic part of
$b(\dS)$.  This term is absent in the Taylor series approximation proposed
in eq.~(7) of ST02.  The latter is obtained by replacing the barrier
$\mathcal{B}$ and the variance $S$ in their unconditional mass function in
the ellipsoidal model with
$\mathcal{B}(S(M_1),z_1)-\mathcal{B}(S(M_0),z_0)$ and $S(M_1)-S(M_0)$,
respectively; that is, they assume that the unconditional and conditional
mass functions have the same form.  This assumption holds exactly in the
spherical case but is not so for ellipsoidal collapse.  As we illustrate in
Fig.~1 below, their expression for the conditional mass function becomes
inaccurate for small look-back times.

We can now use eq.~(\ref{dNdM_f}) to convert our first-crossing
distribution in eq.~(\ref{f_complete}) into the conditional mass function.
We first need to specify the coefficients $b_0, b_1$, and $b_2$ for the
barrier $b(\dS)=\mathcal{B}[S(M_1),z_1]-\mathcal{B}[S(M_0),z_0]$.  To do
so, we use the barrier shape from \cite{SMT01} and ST02 that has been shown
to provide close fits for the {\it unconditional} mass function:
\begin{equation}
\label{Barrier}
     \mathcal{B}[S(M),z]=\sqrt{\gamma}\omega(z)\left[1+\beta(\gamma\nu)^{-\alpha}\right]
\end{equation} 
where $\alpha=0.615$, $\beta=0.485$, $\gamma=0.75$, $\nu=\omega^2(z)/S(M)$,
$\omega(z)=\delta_c/D(z)$, $\delta_c=1.68$, and $D(z)$ is the linear growth
factor. When $z_1-z_0 \ll 1$, it is straightforward to show that
$b_0=\Delta\omega A_0$, $b_1=A_1/\sqrt{S(M_0)}$, and $b_2=-4A_2/[2\pi
  S^3(M_0)]^{1/2}$, where $A_0=0.866(1-0.133\nu_0^{-0.615})$,
$A_1=0.308\nu_0^{-0.115}$, $A_2=0.0373\nu_0^{-0.115}$,
$\nu_0=\omega^2(z_0)/S(M_0)$, and $\Delta\omega=\omega(z_1)-\omega(z_0)$.
The resulting conditional mass function in our ellipsoidal collapse model
for $z_1-z_0 \ll 1$ is:
\begin{eqnarray}
\label{NM}
  && \frac{d}{dM_1}N(M_1,z_1|M_0,z_0)=\left.\frac{d}{dM_1}N(M_1,z_1|M_0,z_0)\right
    \vert_{\rm sph}  \\ \nonumber 
  &\times & A_0\exp\left(-\frac{A_1^2\tilde{S}}{2}\right)
 \left\{1+ A_2\tilde{S}^{3/2}\left[1+\frac{A_1\tilde{S}^{1/2}}{\Gamma(3/2)}\right]\right\}
\end{eqnarray}
where $dN/dM_1\vert_{\rm sph}=(M_0/M_1) (dS_1/dM_1) (\Delta\omega/\Delta S
\sqrt{2\pi\Delta S})$ is the standard spherical model result, $A_0, A_1$,
and $A_2$ are related to $\nu_0$ defined above, and $\tilde{S}=\Delta
S/S(M_0)$.  We find that neglecting the $A_2$ term in eq.(\ref{NM}) (or
$b_2$ in eq.~(\ref{f_complete})) leads to a systematic error of $\gsim
20\%$ for $M_1\sim 0.01M_0$.  It is also worth noting that since the
barriers at different redshifts can intersect in the moving barrier model
(see appendix A of ST02 for more details), $A_0$ in eq.~(\ref{NM}) can be
negative. This occurs very rarely, however, since $A_0<0$ only when $S(M_0)\gsim
30\omega^2(z_0)$, \ie when the descendant mass is much smaller than the
typical halo mass at $z_0$.

We can now use eqs.~(\ref{relation}) and (\ref{NM}) to write down an
analytic expression for the halo merger rate.  Following the notation of
\cite{FM07}, we express the merger rate in terms of the total descendant
mass $M_0=M_1+M_2$ and the mass ratio of the two progenitors $\xi=M_2/M_1$
(assuming $\xi\leq 1$), and use
$B(M_0,\xi,z)dM_0d\xi=B_M(M_1,M_2,z)dM_1dM_2$ to relate the two rates.  By
treating the average merger rate {\it per descendant halo}, $B(M_0, \xi,
z)/n(M_0,z)$, as a single physical quantity, we find
\begin{eqnarray}
\label{B0}
    &&\frac{B(M_0, \xi, z)}{n(M_0,z)}=\left.\frac{B(M_0, \xi, z)}{n(M_0,z)}\right
        \vert_{\rm sph}\\ \nonumber
       &\times& A_0\exp\left(-\frac{A_1^2\tilde{S}_i}{2}\right)
        \left\{ 1+A_2\tilde{S}_i^{3/2}\left[1+\frac{A_1\tilde{S}_i^{1/2}}
       {\Gamma(3/2)}\right]\right\} 
\end{eqnarray}
$\tilde{S}_i=\Delta S_i/S(M_0)$, $\Delta S_i=S(M_i)-S(M_0)$, and the
prediction from the spherical collapse model is
\begin{equation}
\label{B0_spherical}
    \left.\frac{B(M_0, \xi, z)}{n(M_0,z)} \right\vert_{\rm sph} = 
    \frac{d\omega}{dz}\frac{M_0^2}{(1+\xi)^2M_i}\frac{dS(M_i)}{dM_i}
       \frac{1}{\Delta S_i\sqrt{2\pi\Delta S_i}} 
\end{equation}
where $M_i$ can be either of the progenitors $M_1$ or $M_2$. We recall that
the conditional mass function and the merger rate in the EPS model is not
symmetric with respect to the two progenitor masses.  This remains an
unsolved problem. Below we simply show the results for both choices.
According to the notation of \cite{FM07}, we call the merger rate ``option
A'' when $M_i$ is assigned to the smaller progenitor $M_2$, and ``option
B'' when $M_i=M_1$.

\section{Numerical Results and Comparison with the Millennium Simulation}
\label{Millennium}

\begin{figure}
\setlength{\epsfxsize}{0.5\textwidth}
\centerline{\epsfbox{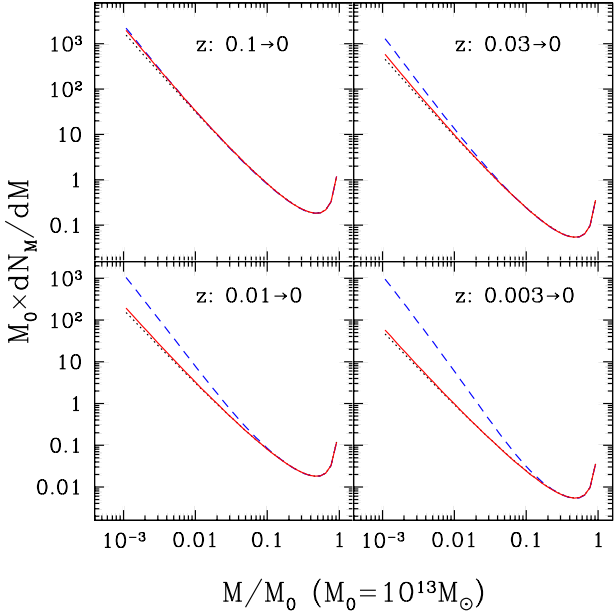}}
\caption{The conditional mass functions for the progenitor halos of a
  descendant halo of mass $M_0=10^{13}M_{\odot}$ at $z=0$. The four
  panels are for four look-back times: $\Delta z=0.1, 0.03, 0.01$, and
  0.003.  Eq.~(\ref{NM}) of this paper (red solid) agrees closely
  with the exact solution from the method of ZH06 (black dotted), while the
  approximation based on eq.~(7) of ST02 (blue dashed) overpredicts the
  number of progenitors for small look-back time ($\Delta z \lsim 0.03$).
}
\label{NM_compare}
\end{figure}

In Fig.~\ref{NM_compare}, we illustrate the accuracy of eq.~(\ref{NM}) by
comparing it with the exact solution from ZH06 and the analytic
approximation based on eq.~(7) of ST02. The figure shows the
conditional mass functions at four different look-back times ($\Delta
z=0.1, 0.03, 0.01, 0.003$) for a descendant halo of mass $10^{13}M_{\odot}$
at redshift zero.  The approximation of ST02 is seen to overpredict
the number of lower mass progenitors by up to a factor of 2 to 10 for
$\Delta z \lsim 0.03$, whereas eq.~(\ref{NM}) of this paper is accurate
when $\Delta z$ is as small as 0.003.

For the halo merger rate, we compare the predictions of our ellipsoidal model
with the Millennium simulation merger rate determined by \cite{FM07} using the
``stitching'' method.  They find that the halo merger rate in the
simulation converges well when the look-back time $\Delta z$ approaches
zero and can be described by a simple universal fitting formula:
\begin{equation}
\label{B_Mi}
     \frac{B(M_0, \xi, z)}{n(M_0,z)} = A\left(\frac{M_0}{\tilde{M}}\right)^{\alpha_1}\xi^{\alpha_2}\exp\left[\left(\frac{\xi}{\tilde{\xi}}\right)^{\alpha_3}\right]
         \left[\frac{d\omega(z)}{dz}\right]^{\alpha_4}
\end{equation} 
where $\tilde{M}=1.2\times 10^{12}M_{\odot}$, $\tilde{\xi}=0.098$,
$A=0.0289$, $\alpha_1=0.083$, $\alpha_2=-2.01$,
$\alpha_3=0.409$, and $\alpha_4=0.371$.  Their Fig.~15 illustrates the
large discrepancy between eq.~(\ref{B_Mi}) and the prediction of the
standard spherical EPS model.

\begin{figure}
\setlength{\epsfxsize}{0.5\textwidth}
\centerline{\epsfbox{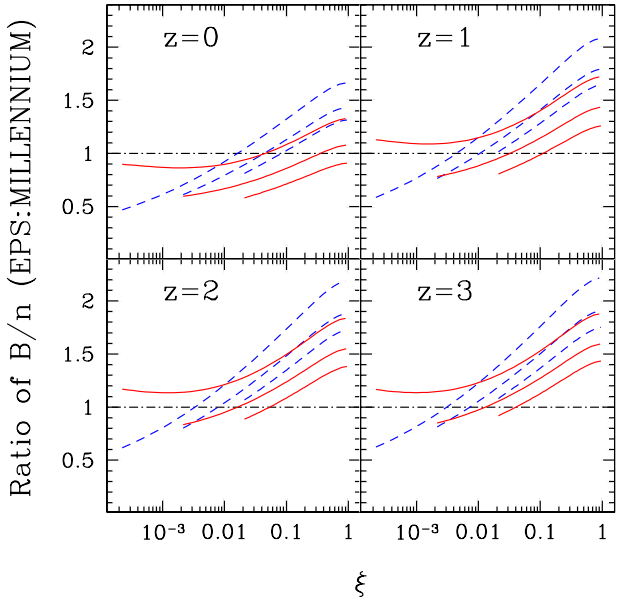}}
\caption{Comparison of the halo merger rates from our ellipsoidal model
  predictions (eq.(\ref{B0},\ref{B0_spherical})) vs. the Millennium
  simulation (eq.(\ref{B_Mi})).  The four panels show the merger rates as a
  function of the progenitor mass ratio $\xi \equiv M_2/M_1$ at $z=0, 1, 2,
  3$.  Within each panel, the red solid curves show the ratio of our
  ellipsoidal collapse model prediction (eq.(\ref{B0})) to the Millennium
  result; the blue dashed curves show the ratio of the standard spherical
  collapse model (eq.(\ref{B0_spherical})) to the Millennium result.  For each
  colour, the set of three curves show three descendant halo masses (from
  bottom up): $10^{12}$, $10^{13}$, and $10^{14}M_{\odot}$.  The progenitor
  mass $M_i$ in eqs.~(\ref{B0_spherical}) and (\ref{B0}) is chosen to be
  the less massive $M_2$ (option A), which we find to match the simulation
  better than option B.}
\label{mr1}
\end{figure}

\begin{figure}
\setlength{\epsfxsize}{0.5\textwidth}
\centerline{\epsfbox{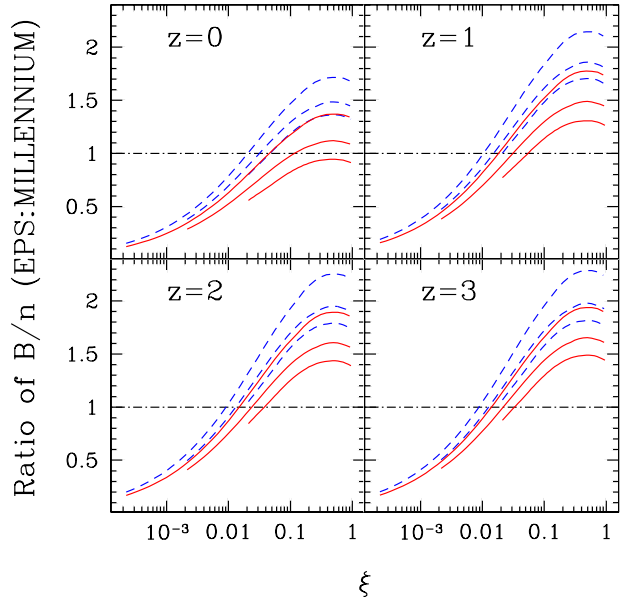}}
\caption{Same as Fig.~\ref{mr1}, except that the progenitor mass $M_i$ in
  both eq.~(\ref{B0_spherical}) and (\ref{B0}) is chosen to be the more
  massive progenitor $M_1$ (option B), which is not our preferred option.}
\label{mr2}
\end{figure}

Figs.~\ref{mr1} (for option A) and \ref{mr2} (option B) show the ratio
between the halo merger rate $B/n$ of our ellipsoidal collapse model
(eq.~(\ref{B0})) and that of the Millennium simulation (eq.(\ref{B_Mi}))
for three descendant halo masses and four redshifts.  The spherical
collapse model (eq.~(\ref{B0_spherical})) is also shown for comparison.
The minimum halo mass is chosen to be $2\times 10^{10}M_{\odot}$ as
set by the halo mass resolution in the Millennium simulation.  Comparison of
the two figures shows that the two choices of $M_i$ give similar results
for major mergers but yield very different predictions for $\xi\ll 1$,
where option A (Fig.~\ref{mr1}) agrees better with the Millennium than
option B (Fig.~\ref{mr2}).  We note that the two options predict different
power-law dependencies on $\xi$ at $\xi\ll 1$: for $S(M)\propto
M^{-\gamma}$, option A gives $B/n\propto\xi^{\gamma/2-2}$, but option B
gives $B/n\propto\xi^{-1.5}$, which is independent of the density variance
on the scale of the smaller progenitor mass.  It is also interesting to
note that option A is implicitly used in some Monte Carlo methods; for
example, \cite{CLBF00} select the mass of the first progenitor from the
lower half of the conditional mass function (i.e. $M_i < M_0/2$).

According to Fig.~\ref{mr1}, the discrepancy between our ellipsoidal
collapse model and the Millennium simulation is typically $20\sim 30\%$,
but can reach up to about $80\%$ when $\xi\sim 1$.  On the other hand, the
relative difference between the spherical collapse model and Millennium is
typically $40\% \sim 60\%$, and reaches up to $120\%$ for the major
mergers.  Therefore in almost every case, the new ellipsoidal collapse
model improves the agreement with the Millennium simulation.

\section{Summary}
\label{summary}

We have derived new analytic formulae for the first-crossing distribution
(eq.(\ref{f_complete})), the conditional mass function (eq.(\ref{NM})), and
the merger rate of dark matter halos (eq.(\ref{B0})) in the ellipsoidal
collapse model in the limit of small look-back times.  Our method is based
on solving the first-crossing distribution of random walks with a moving
barrier using the exact integral equation of ZH06.  This method results in
extra terms in eqs.~(\ref{f_complete}), (\ref{NM}), and (\ref{B0}) that are
absent in the spherical collapse model and different from those in the
ellipsoidal model of ST02.  Fig.~1 illustrates how these terms correct the
discrepancies of ST02 in the conditional mass function for small look-back
times $\Delta z\la 0.03$.

Eq.~(\ref{relation}) relates the conditional mass function at small
$\Delta z$ to the halo merger rate.  The halo merger rate of our
ellipsoidal collapse model generally agrees better with the Millennium
result reported in \cite{FM07} than that of the spherical collapse model
(red vs blue curves in Figs.~\ref{mr1} and \ref{mr2}).  The discrepancy
between our ellipsoidal collapse model and the Millennium simulation is
typically $20\% \sim 30\%$, which is about a factor of two smaller than
that of the spherical collapse model.  A comparison between Figs.~\ref{mr1}
and \ref{mr2} shows a better agreement between model and simulation when
the progenitor mass $M_i$ in the analytic formulae is assigned to be the
smaller progenitor (option A).

A number of factors not considered in this paper nor in earlier EPS work
can contribute to the remaining 20-30\% discrepancy between the model
and simulation.  These include the statistical importance of non-binary
mergers, diffuse accretion and tidal stripping of halo mass (such that
$M_0\neq M_1+M_2$), and the impact of a halo's environment on merger rates
that can lead to non-Markovian processes in the excursion set model (\eg
ST02, \citealt{ND07}).

Our conditional mass function in eq.(\ref{NM}) can be easily incorporated
into Monte Carlo simulations to study halo merger histories over a large
look-back time.  This is done in a companion paper, in
which we compare several existing Monte Carlo algorithms (\eg,
\citealt{LC93,KW93,SK99,CLBF00}) that are all based on the spherical
collapse model, and propose a more accurate method using eq.(\ref{NM}).
Several groups have recently proposed accurately parameterized forms of the
conditional mass function based on fits to the Millennium results and
incorporated them into different Monte Carlo simulations
(\citealt{Cole07,Parkinson07,ND07}).  As we will show in the companion paper (also
see \citealt{MS07} for a different method), a similar level of accuracy can
be achieved using eq.(\ref{NM}) of this paper and our Monte Carlo method
without a priori knowledge of N-body simulations.

We thank James Bullock, Joanne Cohn, Lam Hui, Ravi Sheth, Martin White, and
Simon White for useful discussions.  This work is supported in part by NSF
grant AST 0407351.  The Millennium Simulation databases used in this paper
and the web application providing online access to them were constructed as
part of the activities of the German Astrophysical Virtual Observatory.

\bibliographystyle{mn2e}

\label{lastpage}
\end{document}